\documentclass[12pt]{article}
\usepackage{amsmath}
\usepackage{graphicx,psfrag,epsf}
\usepackage{enumerate}
\usepackage[numbers]{natbib}
\usepackage{amssymb}
\usepackage{caption}
\usepackage{multirow}
\usepackage{xcolor}
\usepackage{subcaption}
\usepackage{setspace}
\usepackage{url} 

\newcommand{\blind}{1}
\begin{document}

\bibliographystyle{plainnat}

\def\spacingset#1{\renewcommand{\baselinestretch}%
{#1}\small\normalsize} \spacingset{1}


\if1\blind
{
  \title{\bf Walking fingerprinting}
  \author{
   Lily Koffman\thanks{
    Correspondence to lkoffma2@jh.edu}\hspace{.2cm}\\
    \small{Department of Biostatistics, Johns Hopkins Bloomberg School of Public Health}\\
    Ciprian Crainiceanu \\
    \small{Department of Biostatistics, Johns Hopkins Bloomberg School of Public Health} \\
    Andrew Leroux \\
    \small{Department of Biostatistics and Informatics, Colorado School of Public Health}
    }
  \maketitle
} \fi

\if0\blind
{
  \bigskip
  \bigskip
  \bigskip
  \begin{center}
    {\LARGE\bf Walking fingerprinting}
\end{center}
  \medskip
} \fi

\medskip
\begin{abstract}

We consider the problem of predicting an individual's identity from accelerometry data collected during walking. In a previous paper we introduced an approach that transforms the accelerometry time series into an image by constructing its complete empirical autocorrelation distribution. Predictors derived by partitioning this image into grid cells were used in logistic regression to predict individuals. Here we: (1) implement machine learning methods for prediction using the grid cell-derived predictors; (2) derive inferential methods to screen for the most predictive grid cells; and (3) develop a novel multivariate functional regression model that avoids partitioning of the predictor space into cells. Prediction methods are compared on two open source data sets: (1) accelerometry data collected from $32$ individuals walking on a $1.06$ kilometer path; and (2) accelerometry data collected from six repetitions of walking on a $20$ meter path on two separate occasions at least one week apart for $153$ study participants. In the $32$-individual study, all methods achieve at least $95$\% rank-1 accuracy, while in the $153$-individual study, accuracy varies from $41$\% to $98$\%, depending on the method and prediction task. Methods provide insights into why some individuals are easier to predict than others. 
\end{abstract}

\vfill

\newpage
\spacingset{1.45} 
\section{Introduction}
\label{sec:intro}

The study of human gait has long been an area of interest for research. Since perhaps the first published study on gait - Aristotle's ``On The Gait of Animals" \cite{connor_biometric_2018} - major strides have been made in collection and analysis of gait data. Various theories have been proposed to explain the  process of human walking, from game-theory based mathematical models \cite{iosa_golden_2017} to dynamic kinematic principles \cite{kuo_dynamic_2010}. Recent technological advances have facilitated the collection of massive amounts of walking data from devices including video cameras, force plates, accelerometers, and gyroscopes. 

A single step is a complex coordinated movement. An individual's gait is a function of stride length, step cadence, joint angles, foot shape, and center of mass, all of which are influenced by height, weight, body composition, and fitness. Gait can vary with time of day, fatigue, emotional state, aging and disease processes. Nonetheless, gait has been shown to be similar enough within individuals, different enough between individuals, and hard enough to permanently change or mimic \cite{gafurov_spoof_2007} that it can be used for identification \cite{connor_biometric_2018}.

Gait-based identification, if achieved with high accuracy, has many promising applications in the fields of biometrics, medicine, and epidemiology. Like fingerprints, retinal scans, or face recognition, gait may be used for identity confirmation. It may even be preferable to other methods because re-authentication can be performed continuously and its use may alleviate  privacy concerns associated with the storage of fingerprints or photos \cite{vildjiounaite_unobtrusive_2006}. Health status and age may be associated with gait patterns or changes in gait \cite{cohen_gait_2019, samson_differences_2001}; as such, in clinical and epidemiological settings, characterizing gait at a specific point in time and quantifying deviations from this baseline may provide a mechanism to measure disease progression or recovery from an adverse health event such as stroke. Changes in gait have also been shown to predict increased fall risk \cite{maki_gait_1997} or undetected disease \cite{hausdorff_gait_2009}.

While the potential benefits of gait-based identification are numerous, it is a difficult task. Methods for gait-based identification are sensitive to an individual changing clothes, shoes, or walking surface \cite{sarkar_humanid_2005}. Collection of high quality walking data is challenging, few open-source data sets exist, and walking in a controlled setting may differ from walking in the real world. As a result, gait has not been widely implemented as a biometric signature, and methods for gait-based identification are not frequently used in epidemiological research or clinical practice.

In this paper, we focus on gait-based identification from high resolution accelerometry data. Compared to video and underfoot-force data, the use of accelerometery in gait recognition and quantification is relatively new.  However, in recent years, the use of accelerometers in physical activity research has proliferated \cite{karas_accelerometry_2019}. This is likely due to the wide social acceptance and ubiquity of wearable devices, substantial improvements in battery life and technology, and the convenience of continuous use of these devices during daily living activities. Therefore, collecting accelerometry data during walking is feasible, un-intrusive, and provides information during a wide range of physical activities that are part of an individual's life.  Increasingly, acceleration during walking is viewed as a technology with extraordinary potential for gait-based identification and quantification \cite{gafurov_biometric_2006}.

Existing methods for gait-based identification from accelerometry data can be divided into those that rely on step cycle detection or stride segmentation and those that are step cycle independent. Typically, methods that rely on step cycle detection segment the accelerometry pattern into steps, average over many steps to create a subject-specific template, and then predict unidentified data by matching to these templates based on cross-correlation or other distance metrics. An early implementation of this approach achieved $7$\% equal error rate (EER) in a study of $36$ individuals \cite{mantyjarvi_identifying_2005}. Subsequent approaches used variations of cycle matching, including implementing dynamic time warping to normalize step length \cite{rong_wearable_2007} or match cycles \cite{derawi_unobtrusive_2010},  matching with Euclidian distance instead of cross-correlation \cite{gafurov_improved_2010}, and matching on the principal components of the step cycle \cite{bours_eigensteps_2010}; other summaries of the step cycle including Fourier coefficients and histogram features have also been considered \cite{mantyjarvi_identifying_2005, gafurov_biometric_2006}. While methods based on step cycle detection have achieved low error rates, cycle detection can be error prone, computationally intensive, and sensitive to movements of the device, especially on the wrist. Nonetheless, the majority of existing methods for identification from accelerometry data rely on step cycle detection. Notable exceptions include the signature points approach  \cite{zhang_accelerometer-based_2015}, the hidden Markov model approach \cite{nickel_using_2011}, and our own walking fingerprinting \cite{koffman_fingerprinting_2023}. Few of these existing methods are applied in open-source data sets or provide validated software for implementation.

Previously, we have proposed a walking fingerprinting approach \cite{koffman_fingerprinting_2023} using the transformation of the accelerometry time series into an image by considering the complete autocorrelation distribution. This distribution was partitioned into grid cells, predictors were derived by summing over cells, and a logistic model on these cells was used to predict individuals. Here we: (1) expand the grid cell-based approach to include machine learning methods, (2) derive inferential methods to screen for the most predictive grid cells, and (3) develop a novel multivariate functional regression model that avoids partitioning of the predictor space into cells. Methods are compared on two open source data sets that comprise different populations and data collection settings. The first data set contains $32$ individuals with long walking periods and was used in our original paper \cite{koffman_fingerprinting_2023}. The second data set contains $153$ individuals with shorter walking periods collected in two separate sessions, sometimes weeks apart. 

The paper is organized as follows: Section 2 describes how the data were collected and demonstrates examples of the data structure. Section 3 describes methods for identity prediction and inference. Section 4 describes the results of the applications of the methods in the two separate data sets. Discussion and future directions are covered in section 5.

\section{Data Description}\label{sec:data_description}

\begin{figure}[!ht]
    \centering
    \includegraphics[width = 0.75\textwidth]{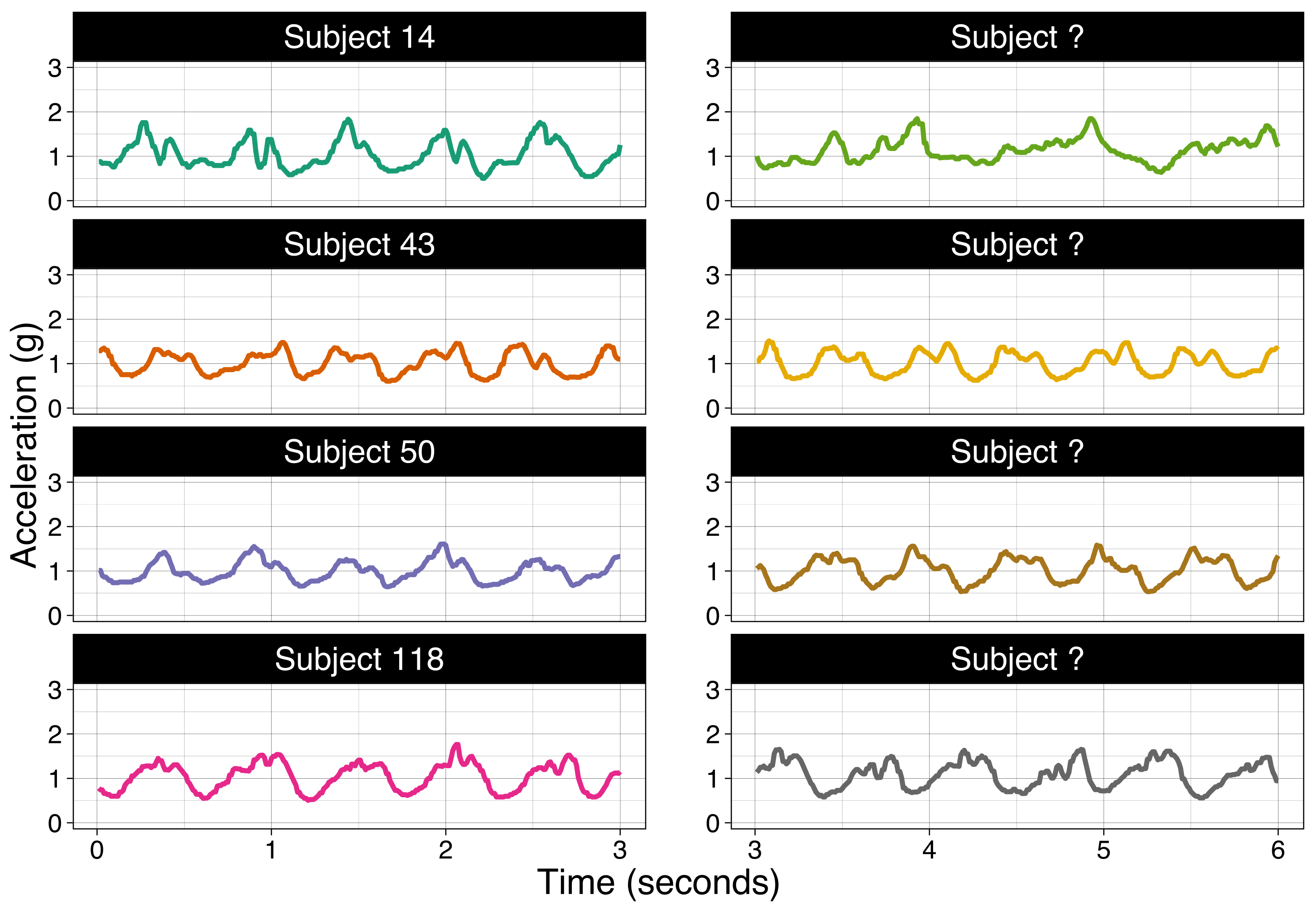}
    \caption{Eight three-second intervals shown from different study participants. Data are a single time series obtained as the sum of squares of observed accelerations along the three axes. The left panels provide the information about the identity of study participants, whereas the right panels do not. The questions are: (1) among the individuals shown in the right plots, is there any individual whose data are displayed in the left panels? and (2) if yes, then which ones?}
    \label{fig:problem_description}
\end{figure}

Wearable accelerometers contain a micro-electromechanical system that records acceleration along three orthogonal axes in the frame of reference of the device. The frequency of most wearable accelerometers ranges from $10$ to $200$ hertz (Hz) ($10$ to $200$ observations per second). Thus, the raw accelerometry data obtained from these wearable devices consists of three simultaneous time series with between $10$ and $200$ observations per second \cite{karas_accelerometry_2019}. In practice, the sum of squared acceleration in each dimension is often used to minimize the effect of small movements of the device around the wrist.  Figure \ref{fig:problem_description} displays the sum of squared acceleration in eight walking intervals from several individuals. Each panel corresponds to a three second interval and the left panels show that data was obtained from four different individuals (labeled $14$, $43$, $50$, and $118$, respectively). The right panels do not show the identity of the individuals and the questions are: (1) among the individuals shown in the right panels, is there any individual whose data are shown in the left panels? and (2) if yes, then which ones? Now, imagine the same problem but with thousands or million of individuals and with varying lengths of walking intervals.  Here we try to address such problems, but using computers, accurately, and fast. Do not worry, we will solve this puzzle for you at the end of the paper. 

\subsection{Indiana University Biostatistics dataset}\label{subsec:IUB}
As part of a study to explore the accelerometry patterns associated with various activities, $32$ participants ($13$ male, $23$ to $54$ years old) wore four ActiGraph GT3X+ accelerometers on the left hip, left wrist, and both ankles, respectively, while walking, driving, and climbing and descending stairs. The study was conducted at the Department of Biostatistics Fairbanks School of Public Health at Indiana University and all subjects provided written informed consent. The data were collected at $100$ Hz and extracted using ActiLife software version 6.12.0. For our purposes, only data from the left wrist collected during the walking part of the study is used. Thirty-one subjects identified as right-handed and one identified as ambidextrous; as such, the left-wrist was effectively the non-dominant wrist for all individuals. 
The walking portion of the study consisted of $1.06$ kilometers of walking on a paved, outdoor path at a comfortable, self-selected pace. To guarantee data quality, participants were instructed to clap three times at the start and end of each activity and the large spikes in acceleration were used to label each activity. To create smooth transitions between activities, $0.5$ seconds before and after the start of each activity were deleted. The total walking time ranged from $6$ minutes to $10$ minutes across participants \cite{fadel_differentiating_2019}.  The raw data are publicly available at \url{physionet.org/content/accelerometry-walk-climb-drive/1.0.0/}
\subsection{Zhejiang University (ZJU) GaitAcc dataset}

The ZJU-GaitAcc dataset consists of data collected while walking for the purpose of gait-based identification. One hundred and seventy five volunteers (approximately $2/3$ male, aged $16$ to $40$ years old) wore five Wii remotes containing an ADXL330 triaxial accelerometer placed at the left upper arm, left thigh, right wrist, right pelvis, and right ankle. The choice of side for placement of the devices was the same for all subjects; the authors claimed that wear side was not particularly important due to the symmetry of gait; this is debatable, but not something under our control. Data were  measured at at least $\pm$ $3$g (standard Earth gravitational unit) and typically up to $\pm$ $5$g with the approximate precision of $(5/128)$g and transmitted via Bluetooth at up to $100$ Hz. The signal was received by a computer and resampled at exactly $100$ Hz. The trial consisted of walking at a comfortable, self-selected pace along a $20$ meter (m) flat path six times in a row. Each repetition was seven to 15 seconds in length and consisted of seven to 14 steps. Portions of walking were manually labelled, and data corrupted due to bad wireless connection between the Wii and the computer were manually removed. In contrast to other existing gait acceleration data sets, the ZJU-GaitAcc data includes a second walking session between one week and six months after the initial session. One hundred and fifty three subjects were present for both sessions. Volunteers were not given any instructions on clothing or footwear for either session except they were asked to not wear slippers. As such, the second session presents a unique opportunity to test some of the challenges associated with gait prediction. Indeed, individuals are likely to change clothing, footwear, and accelerometer placement, while their physical and emotional states may also be different between the two sessions \cite{zhang_accelerometer-based_2015}. For our purposes, we use only the $153$ subjects who were present for both sessions in the following analyses. The data are publicly available for non-commercial use at \url{www.ytzhang.net/datasets/zju-gaitacc}

\section{Methods for Subject Identification}
\label{sec:meth}
\subsection{Notation}
Let $x_{ij}(s), y_{ij}(s), z_{ij}(s)$ denote acceleration measured in Earth gravitational units (g=$9.81 {\rm m/s}^2$) along three orthogonal axes for subject $i$, $i = 1, \dots, N$, at second $j$, $j = 1, \dots, J_i$, and centisecond (one one hundredth of a second) $s$, $s = 1,\dots, S=100$ (since in both datasets, observations are collected at $100$ Hz). Here $J_i$ depends on study participants, because the number of seconds of data is different across individuals. We have two indicators for time, one that counts the seconds, $j$, and one that counts the centiseconds within each second, $s$, because our basic unit of analysis is the second and analytic methods are based on the accelerometry patterns in a collection of seconds for each individual. Each second contains $100$ sampling points for a total of $300$ observations along the three axes.  For each observation, we calculate the vector magnitude of acceleration $v_{ij}(s) = \sqrt{x^2_{ij}(s)+ y^2_{ij}(s) + z^2_{ij}(s)} $, which is rotation-invariant and less sensitive to small movements or location changes of the accelerometer.

For the purposes of this paper the data for subject $i$ consists of $v_{ij}(s)$ $j=1, \ldots, J_i$, $s =1,\ldots,  S=100$ for a total of $100\cdot J_i$ observations.  Indeed, some individuals have less observed walking time simply because they walked faster. Given a sub-sample of these observed walking data (training data) from subjects $i = 1, \dots, N$, our goal is to build models that can identify individuals from their remaining data (testing data). We employ two separate frameworks: prediction using covariates derived from the transformed time series, which we refer to as ``image partitioning," and functional regression, which avoids partitioning of the predictor space. For both approaches, we first obtain the complete autocorrelation distribution of the time series, which can be represented as an image. For the image partitioning approach, we compute summary measures of a subset of the complete empirical autocorrelation distribution and then use these summaries to predict individuals. The functional approach avoids image partitioning and uses the entire autocorrelation distribution in a functional model. We first describe obtaining the autocorrelation distribution, then describe the two modeling approaches. 

\subsection{Obtaining the autocorrelation distribution}\label{subsec:ADF}

To transform the raw time series, we first choose an interval length $S$ and segment the time series into $J_i$ non-overlapping intervals of length $S$ for each subject. In our case we will use $S=100$ centiseconds (equal to a one second interval), but other choices may be more appropriate in other time series applications.

Next, for each time lag $u =1,\ldots, S-1$ we construct the set of three dimensional vectors $\{v_{ij}(s-u), v_{ij}(s), u\}$ for $s = u+1, \dots, S$. The first entry in this vector is the the lagged time series (observed at $s-u$), the second entry is the time series (observed at $s$), and the last entry is the lag (denoted by $u$). For a fixed lag $u$ the number of vectors $\{v_{ij}(s-u), v_{ij}(s), u\}$ is equal to  $S-u$, one for each $s=u+1,\ldots,S$. For example, for $u = 1$ (equivalent to a $0.01$ seconds lag), there are $S-u = 100-1 = 99$ such vectors for every $i=1,\ldots, N$ and $j \in 1,\ldots, J_i$: 
$\{v_{ij}(1), v_{ij}(2), 1\}, \{v_{ij}(2), v_{ij}(3), 1\}, \dots, \{v_{ij}(99), v_{ij}(100), 1\}\;.$
For $u = 99$ (equivalent to a $0.99$ seconds lag), there is only one vector: $S-u = 100-99 = 1$ for every $i=1,\ldots, N$ and $j \in 1,\ldots, J_i: \{v_{ij}(1), v_{ij}(100), 99\}\;.$

For each subject, $i$, the collection of three dimensional vectors $\{v_{ij}(s-u), v_{ij}(s), u\}$ for all intervals $1, \dots, J_i$ and all lags $u=1,\ldots,S-1$ is a three-dimensional image representation of the vector magnitude time series. We refer to this as the complete autocorrelation distribution of the time series. The number of observations for subject $i$ in this image is equal to 
$\sum_{u=1}^{S-1}(S-u)J_i=J_i\sum_{u=1}^{S-1}u=J_i\frac{S(S-1)}{2}\;.$

\begin{figure}[!ht]
    \centering
    \includegraphics[width=.6\textwidth]{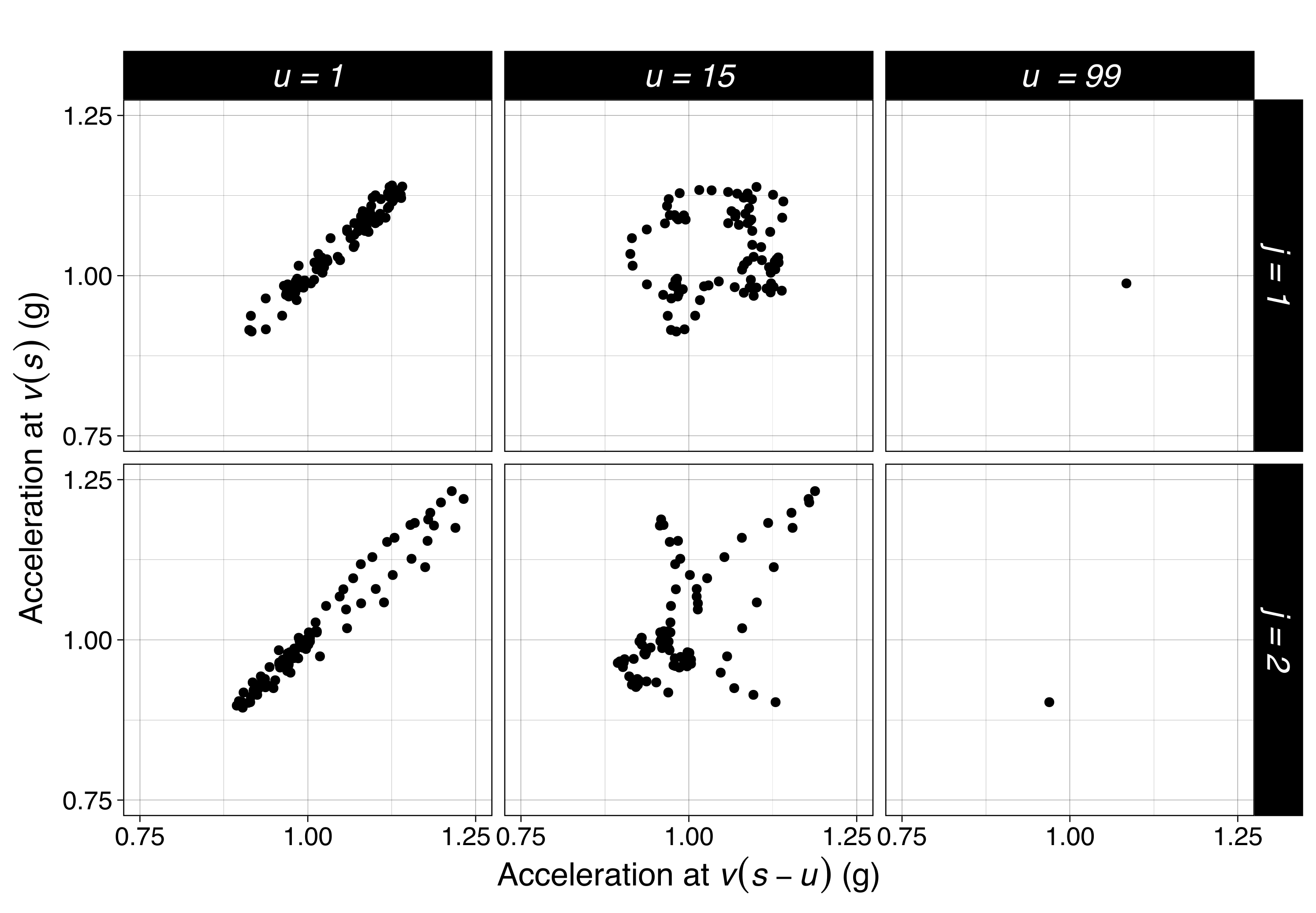}
    \caption{Subset of complete empirical autocorrelation distribution for subject 19 in the IU data. The pairs $\{v_{ij}(s-u), v_{ij}(s)\}$ are plotted for $u = 1, 15, 99$ centiseconds (columns) and $j = 1, 2$ seconds (rows).  
    }
    \label{fig:autocorr_s19}
\end{figure}

Figure~\ref{fig:autocorr_s19} provides the intuition into how this image is constructed. The first row of three panels corresponds to the first second, $j=1$, while the second row corresponds to the second second, $j=2$, of data for subject 19 in the IU data. The columns correspond to three different lags, $u=1$, $15$, and $99$ centiseconds, respectively. Each panel in the first column contains $99$ pairs of observations $\{v_{ij}(s-1), v_{ij}(s)\}$, for $s=1, \ldots, S-1$. Note the strong correlation between the observations, as the values of acceleration do not change too much for one centisecond. The shape of the resulting point cloud resembles that of a bivariate Normal distribution with high correlation for this time lag.  Each panel in the second column contains $85$ pairs of observations $\{v_{ij}(s-15), v_{ij}(s)\}$, for $s=1, \ldots, S-15$. Compared to the first column, the point clouds exhibit lower correlation and exhibit structure beyond simple linear association. Finally, the panels in the last column contain only one pair of observations $\{v_{ij}(s-99), v_{ij}(s)\}$. The three-dimensional image predictor is the union of all these panels over all the seconds, $J_i$, and all lags, $u$, for one individual. We refer to this image as the complete empirical autocorrelation distribution. 

The autocorrelation distribution is the foundation of both prediction approaches we employ. When we predict subject $i_0$ the outcome is $Y_{ij}^{i_0}$, where $Y_{ij}^{i_0}=1$ if $i=i_0$ and $0$ otherwise, i.e. the outcome is an indicator that data belongs to study participant $i_0$. The predictors are $\{v_{ij}(s-u),v_{ij}(s),u:u\in 1,2,\ldots,S-1; s\in u+1,\ldots,S\}$. 
 Thus, at the conceptual level, we changed the problem of identifying individuals from their high density accelerometry data recorded during walking into a binary regression of the type
\begin{equation}
Y_{ij}^{i_0} \; |\; \{v_{ij}(s-u),v_{ij}(s),u:u\in 1,2,\ldots,S-1; s\in u+1,\ldots,S\}
\label{eq:conceptual}
\end{equation}
where there are $J_{i_0}$ indicators equal to one (data observed from study participant $i_0$) and $\sum_{i=1}^n J_i$ indicators of ones and zeros (data for all study participants). One approach to this problem is to reduce the complexity of the predictor space via ``image partitioning," which  summarizes the three dimensional image into several predictors. The second approach is to consider the model $Y_{ij}^{i_0}\sim {\rm Bernoulli} \{p_{ij}^{i_0}\}$ where probabilities are modeled as a tri-variate functional regression model

\begin{equation}
{\rm logit}\{p_{ij}^{i_0}\}= \int_{s,u} F\{v_{ij}(s-u),v_{ij}(s),u\}dsdu
\label{eq:3Dfunctional}
\end{equation} 
In this case parsimony is controlled by assuming that $F(\cdot,\cdot,\cdot)$ is smooth. In the following section we provide more details for both these approaches.

\subsection{Image Partitioning}\label{sec:imagepartition}
The image partitioning approach consists of: (1) transforming the raw times series into an image, which is the complete empirical autocorrelation distribution of the time series; (2) extracting predictors from the transformed time series; (3) selecting important predictors, and (4) fitting prediction models with these predictors. We have already described step 1 and we now describe the other steps.

\subsubsection{Extracting predictors and variable screening}\label{subsec:define_and_screen}

Recall that the complete empirical autocorrelation distribution is obtained using many observations; one of our goals is to reduce the number of observations to a manageable number of predictors while also maintaining interpretability of results and predictive performance. As in our data sets $>99$\% of all values of $v_{ij}(s)$ are between $0$ and $3$g, we consider a partition of the $[0,3]\times [0,3]$ interval in $\mathbb{R}^2$ into squares of length $0.25$g, resulting in a total of $144$ grid cells for each lag $u$. Each pair $\{v_{ij}(s-u), v_{ij}(s)\}$ belongs to one of these cells, and the few cases that do not are discarded from the data set. For example, $\{v_{ij}(s-u), v_{ij}(s)\} = \{0.2, 0.1\} \in [0, 0.25] \times [0, 0.25]$, which happens to be the first grid cell. The number of points (i.e., pairs) in each grid cell for each lag is computed and the collection of the number of points in each grid cell comprises the set of potential predictors. 

We could use all lags, $u=1,\ldots,S-1=99$, but in a previous study \cite{koffman_fingerprinting_2023} we have shown that using only three lags $u = \{15, 30, 45\}$ is actually enough to maintain the prediction performance of our models. The intuition is that data that are very close tend to be highly correlated and do not provide much additional information. For example, the first column panels in Figure~\ref{fig:autocorr_s19} indicate how correlated $v_{ij}(s-1)$ and $v_{ij}(s)$ are. While using more lags is conceptually and practically possible, this choice provides a good balance between complexity and predictive performance.  

\begin{figure}[!ht]
    \centering
    \includegraphics[width=.7\textwidth]{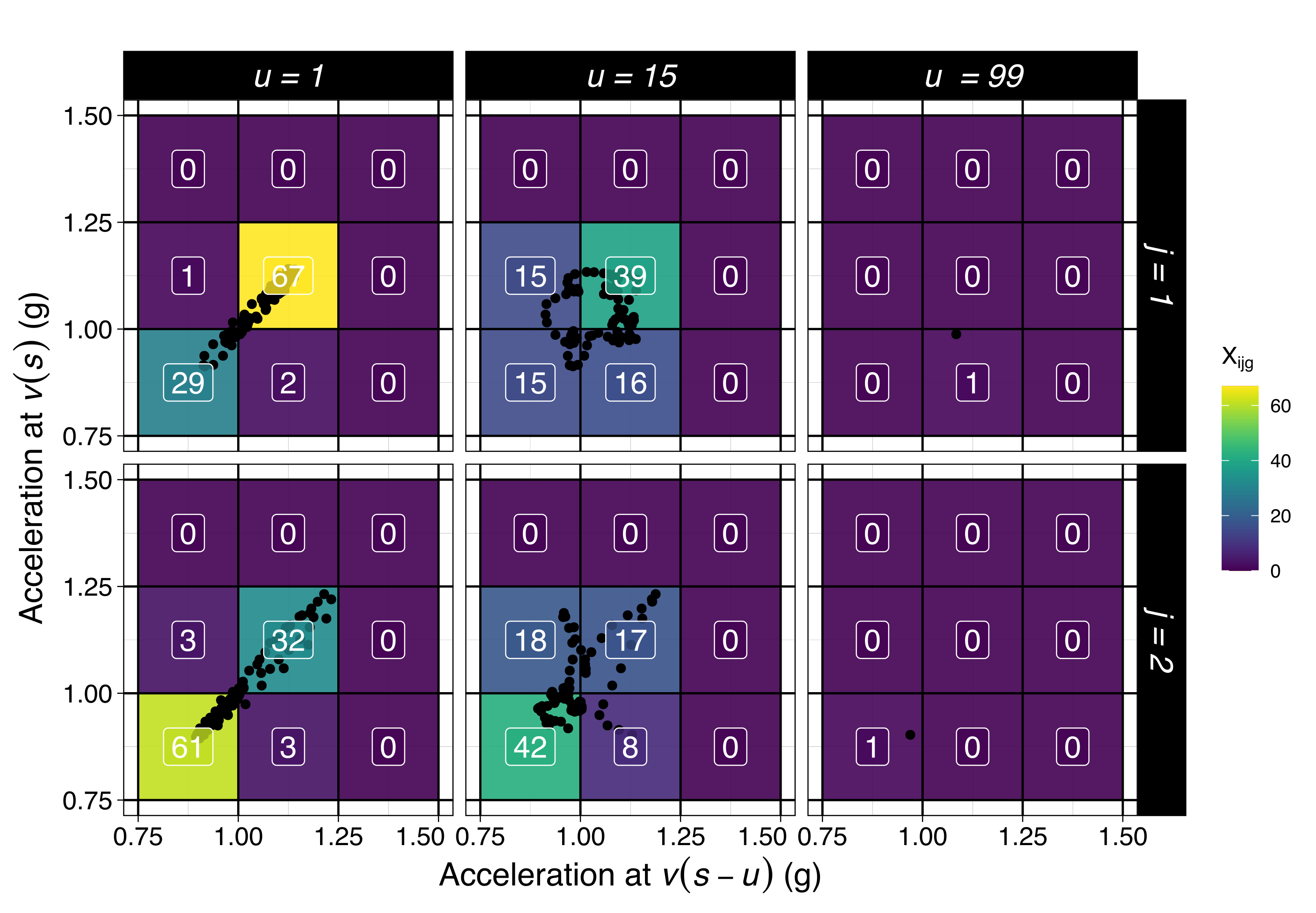}
    \caption{Predictor extraction for subject 19. 
    The values of $X_{ijg}$ for subject 19 in the IU data are shown for $u = 1, 15, 99$ (columns) and $j = 1,2$ (rows). The white number in each grid cell is the value of $X_{ijg}$ for that cell. For example, $X_{i = 19, j = 2, g = [0.75, 1.00), [0.75, 1.00)} = 61$ and is shown in the bottom-left corner of the plot. Only a subset of the grid cells is shown as the other grid cells have no observations for these two seconds} 
    \label{fig:extraction_19}
\end{figure}

With three lags and $144$ cells for each lag  there are a total of $G=3\times 144 =432$ possible cells for each distribution. We define the predictors $X_{ijg}$ as the number of observations $\{v_{ij}(s-u),v_{ij}(u)\}$ for $s=u+1,\ldots,S$ that fall into cell $g$ for subject $i$ and each second $j$.  Figure~\ref{fig:extraction_19} illustrates the process of obtaining $X_{ijg}$ for seconds $1$ (first row) and $2$ (second row) from participant $19$ in the IU data. For illustration purposes, the image is zoomed in on the intervals $[0.75,1.50]\times [0.75,1.50]$ g, which contains only $9$ cells for each lag. However, the original space is $[0.00,3.00]\times [0.00,3.00]$ and contains $144$ cells for each lag.

Let us focus on the panel in the first row and second column, which corresponds to a lag of $u=15$ centiseconds. There are $85$ black dots in this panel corresponding to the pairs $\{v_{i1}(s-15),v_{i1}(s)\}$ for $s=16,\ldots,S=100$. The yellow square (cell) in the middle of the panel contains $39$ of these pairs of observations, which is indicated by the number in the middle of the square and is coded from yellow (high) to purple (zero). For this cell $g$ we build the predictor $X_{i1g}=39$ and the process is repeated for every cell shown in the figure. Thus, for lag $u=15$ we build the predictor vector (enumerating by rows from first row) $(0,0,0,15,39,0,15,16,0)$. Similarly we build predictors based on lags $u=30$ and $u=45$ centiseconds and we simply append all these predictors in a long vector of predictors.

Once these predictors are built, we have transformed the problem described in~\eqref{eq:conceptual} into the regression or prediction problem
\begin{equation}
Y_{ij}^{i_0}\;| \;X_{ij1},\ldots,X_{ijG}
\label{eq:partitioning_reg}
\end{equation}
where $X_{ijg}$ is the number of observations $\{v_{ij}(s-u),v_{ij}(s)\}$ for study participant $i$ in second $j$ and cell $g$, and $Y_{ij}^{i_0}$ is $1$ if $i=i_0$ and $0$ otherwise. The hope is that not much information was lost by summarizing the images via the $X_{ijg}$ variables. The most important consequence of the image partitioning is that it transforms a problem that is intuitively described into a well defined statistical problem. Indeed, if one agrees that this is a good idea, then there are many potential solutions to extract the maximum amount of predictive performance from the $X_{ijg}$ variables.

Because many grid cells contain few observations, we applied a variable screening approach by removing grid cells for which both (1) there are very few unique values; and (2) the ratio of the most common value to the second-most common value is large \cite{recipes}. Specifically, grid cells for which both fewer than $10$\% of values across all seconds and individuals are unique and for which the ratio of the most common to the second most common value is greater than $95:5$ are removed. An example of a hypothetical predictor that would satisfy these criteria is one that, in $100$ samples, has three unique values (say $0$, $1$, and $2$) and the value for $98$ of the $100$ samples is $0$.

\subsubsection{Model Fitting}\label{subsec:modelfitting}
Once the set of predictors is established, they can be used in any classification algorithm. We have already published a method based on logistic regression using all variables obtained from the image partitioning \cite{koffman_fingerprinting_2023}. Here we explore a large number of machine learning models to see whether the performance of logistic regression can be improved. In all models we have employed one versus the rest prediction and we have conducted the same procedure for each study participant; technically, what changes from one model to another is the identifier for the study participant who is predicted at that time.

For the logistic regression approach, a separate multivariable logistic regression model is fit on the training data for each subject. Models provide an estimation of the probability ${\rm Pr}(Y_{ij}^{i_0}=k)$ for every pair of study participants, $(i,k)$, at every second, $j=1,\ldots, J_{\rm test}$. This calculation is done separately for every study participant, $i_0$; recall that a different model is fit for every study participant. To predict the identity of each subject, we first normalize so that all predicted probabilities for each second sum to one, then average these probabilities over all seconds to get a single probability for each subject's potential identity: $$\widehat{\rm Pr}(Y_{i}^{i_0} = k) = \frac{1}{J_\text{test}}\sum_{j=1}^{J_\text{test}}\widehat{\rm Pr}(Y_{ij}^{i_0} = k)\;.$$
Finally, we classify subjects as $\widehat{k}(i_0) = {\rm argmax}_{k} \widehat{\rm Pr}(Y_i=k)$.

For machine learning, several candidate models are tuned using a parameter grid search using  five-fold cross-validation within the training data. The candidate models are: radial basis support vector machine (SVM) \cite{karatzoglou_kernlab_2004}, polynomial basis SVM \cite{karatzoglou_kernlab_2004}, random forest \cite{kuhn_applied_2013}, Bayesian additive regression tree (BART) \cite{chipman_bart_2010}, penalized logistic regression \cite{kuhn_applied_2013}, neural net \cite{kuhn_applied_2013}, boosted tree \cite{chen_xgboost_2016}, naive Bayes classifier \cite{kuhn_applied_2013}, k-nearest neighbors \cite{hechenbichler_weighted_2004}, flexible discriminant analysis \cite{hastie_flexible_1994}, and multivariate adaptive regression splines (MARS) \cite{friedman_multivariate_1991}. All models were fit using tidymodels in R \cite{kuhn_tidy_2023, R, tidymodels}. The model with the best five-fold cross-validated AUC in the training data is then used as the final model and predictions are obtained from fitting this final model on the testing data. The process is repeated for each subject, so different models may be optimal for different study participants. Once the predictions on the test data are obtained, the same one versus the rest classification scheme is used as for the logistic regression. 

\subsubsection{Correlation and multiplicity adjusted (CMA) inference}\label{sec:CMA_inference} 
In logistic regression we can fit the full model and obtain a vector of parameter estimates $\widehat{\boldsymbol{\beta}}$ together with a covariance matrix $\widehat{\mathbf{V}}_{\boldsymbol{\beta}}=\widehat{\rm Var}(\widehat{\boldsymbol{\beta}})$. Confidence intervals are typically obtained as
$\widehat{\boldsymbol{\beta}}\pm z_{1-\alpha/2}{\rm diag}(\widehat{\mathbf{V}}_{\boldsymbol{\beta}})$,
where $z_{1-\alpha/2}$ is the $1-\alpha/2$ quantile of a $\mathcal{N}(0,1)$ distribution and $\text{diag}(\mathbf{A})$ is the diagonal vector of a symmetric matrix $\mathbf{A}$. In our setting, there are two problems with this approach. First, this does not adjust for correlation among tests, which may be large because the $X_{ijg}$ are highly correlated. Second, it does not correct for multiplicity of tests, which can be quite large. To address these issues, we calculate and report correlation and multiplicity (CMA) confidence intervals; for more details, see Chapter 2 in \cite{crainiceanu2023book}. Here we provide a short self-contained summary of the procedure. 

Under the assumption that $\widehat{\boldsymbol{\beta}}$ is multivariate normal, we have $ \widehat{\boldsymbol{\beta}} \sim N(\boldsymbol{\beta}, \widehat{\mathbf{V}}_\beta)$. If
$\mathbf{D}_G = \sqrt{\text{diag}(\widehat{\mathbf{V}}_\beta})$ is the $G\times 1$ dimensional vector of square roots of the diagonal elements of $\widehat{\mathbf{V}}_\beta$, then 
 $(\widehat{\boldsymbol{\beta}}-\boldsymbol{\beta})/\mathbf{D}_G \sim N(\mathbf{0}_G, \mathbf{C}_G)$,
where the ratio of the two vectors is entrywise, $\mathbf{C}_G = \mathbf{V}_G/\mathbf{D}_G \mathbf{D}_G^t$, and the ratio of these two matrices is entrywise. Thus, if we find $q(\mathbf{C}_G, 1-\alpha)$ that satisfies 
$P\{q(\mathbf{C}_G, 1-\alpha)\mathbf{e} \leq (\widehat{\boldsymbol{\beta}}-\boldsymbol{\beta})/{\mathbf{D}_G}  \leq q(\mathbf{C}_G, 1-\alpha)\mathbf{e}\}$,
where $\mathbf{e} = (1, 1, \dots 1)^t$ is a $G\times 1$ dimensional vector of ones, then CMA confidence intervals are: 

$$ \hat{\boldsymbol{\beta}} \pm q(\mathbf{C}_G, 1-\alpha)\sqrt{\text{diag}(\hat{\boldsymbol{V}}_\beta)}\;.$$

We will use the {\ttfamily mvtnorm::qmvnorm} \cite{mvtnorm} function to obtain the quantile $q(\mathbf{C}_G, 1-\alpha)$ using the following one line of code. 
\begin{verbatim} 
 q <- mvtnorm::qmvnorm(p = .95, corr = C,tail = "both.tails")$quantile
\end{verbatim}

This procedure provides an explicit screening approach for finding the walking fingerprint by: (1) regressing the person-specific indicator for each second on all predictors obtained from the image partition; and (2) identifying only those predictors that are significant at a given level $\alpha$ using the CMA procedure.

\subsection{Functional Regression}\label{subsec:functional_regression}
The image partition approach described in Section~\ref{sec:imagepartition} reduces the complexity of images (complete empirical autocorrelation function) and conducts regression in this simplified predictor space. An alternative is to consider the entire image as a predictor, but induce parsimony using smoothing assumptions on the shape of the association between  the image and the probability that the data originates from a particular individual. 
Our functional regression approach starts with the complete data $\{Y_{ij}^{i_0},v_{ij}(s-u),v_{ij}(s),u\}$ for all $i=1,\ldots,N$, $j=1,\ldots, J_i$, $u=1,\ldots,S-1=99$, and $s=u+1,\ldots,S=100$.  Our proposed approach is to fit the following model $Y_{ij}^{i_0}\sim {\rm Bernoulli} (p_{ij}^{i_0})$ where probabilities are modeled as a tri-variate functional regression model

$${\rm logit}(p_{ij}^{i_0}) = \beta_0^{i_0} + \int_{u=1}^{S}\int_{s=u}^S F_{i_0}\{v_{ij}(s-u), v_{ij}(s), u\}dsdu\;,$$
where $F(\cdot,\cdot,\cdot)$ is a trivariate smooth function that takes values at every point in the domain of the three-dimensional images (complete empirical autocorrelation functions). This idea is related to the bivariate functional generalized additive models \cite{cui2021additive,mclean2014functional,mullerGAM}, though the application context here is different, we are working with a trivariate image as a predictor, the domain of the function is not rectangular, and the size of the data sets is much larger than what was previously considered. As we will discuss, we are pushing the boundaries of what these models can actually handle.

The idea is to expand the functional coefficient in a spline basis and penalize the roughness of the function using quadratic penalties. More specifically, we use a Kronecker product of spline basis expansion of the type
$$ \small F(d,v,u)=\sum_{k_d=1}^{K_d}\sum_{k_v=1}^{K_v}\sum_{k_u=1}^{K_u}\beta_{k_d,k_v,k_u}B_{k_d}(d)B_{k_v}(v)B_{k_u}(u)\;,$$
where $B_{k_d}(\cdot)$, $k_d=1,\ldots,K_d$, $B_{k_v}(\cdot)$, $k_v=1,\ldots,K_v$, and $B_{k_u}(\cdot)$, $k_v=1,\ldots,K_v$ are univariate spline bases. We can denote by $\boldsymbol{\beta}$ the $K_dK_vK_u\times 1$ dimensional vector of $\beta_{k_d,k_v,k_u}$ parameters in a specified order.  With this notation the functional regression model becomes

$$ \small {\rm logit}(p_{ij}^{i_0}) = \beta_0^{i_0}+ \int_{u=1}^{S}\int_{s=u}^S \sum_{k_d= 1}^{K_d}\sum_{k_v=1}^{K_v}\sum_{k_u=1}^{K_u}\beta_{k_d,k_v,k_u}B_{k_d}\{v_{ij}(s-u)\}B_{k_v}\{v_{ij}(s)\}B_{k_u}(u)dsdu\;.$$
The double integral in this equation is a theoretical representation of what we would like to calculate. In practice, we approximate the integrals using a Riemann sum. Let $s_l$, $l=1,\ldots,L$ be a grid of points where the integral over $s$ is approximated and $w_l$, $l=1,\ldots,L$ the corresponding Riemann sum weights. Similarly, let $u_m$, $m=1,\ldots,M$ be a grid of points where the integral over $u$ is approximated and $w_m$, $l=1,\ldots,m$ the corresponding Riemann sum weights. With this notation the double integral (and triple sum) can be approximated by the following quintuple-sum:

\begin{align*}
\small
     &=\sum_{m}w_{m}\sum_{l}w_{l}\sum_{k_d= 1}^{K_d}\sum_{k_v=1}^{K_v}\sum_{k_u=1}^{K_u}\beta_{k_d,k_v,k_u} B_{k_d}\{v_{ij}(s_l-u_m)\}B_{k_v}\{v_{ij}(s_{l})\}B_{k_u}(u_{m})\\
&= \sum_{k_d= 1}^{K_d}\sum_{k_v=1}^{K_v}\sum_{k_u=1}^{K_u} \beta_{k_d,k_v,k_u}\sum_{m}w_{m}\sum_{l}w_{l} B_{k_d}\{v_{ij}(s_{l}-u_{m})\}B_{k_v}\{v_{ij}(s_{l})\}B_{k_u}(u_{m})\\
&= \sum_{k_d= 1}^{K_d}\sum_{k_v=1}^{K_v}\sum_{k_u=1}^{K_u} \beta_{k_d,k_v,k_u} C_{ij,k_d,k_v,k_u}\;,
\end{align*}
where $C_{ij,k_d,k_v,k_u}=\sum_{m}w_{m}\sum_{l}w_{l} B_{k_d}\{v_{ij}(s_{l}-u_{m})\}B_{k_v}\{v_{ij}(s_{l})\}B_{k_u}(u_{m})$ are subject and second-specific covariates that correspond to  the $(k_d,k_v,k_u)$ index of the Kronecker product spline basis. One could fit this model directly by calculating the $C_{ij,k_d,k_v,k_u}$ variables and conducting a binary regression of the $Y_{ij}^{i_0}$ outcomes on these covariates.

There are three problems associated with this potential solution. First, the total number of parameters, $K_dK_vK_u$, increases very fast with the number of basis functions in each dimension. For example, if $K_d=K_v=K_u=20$ the total number of predictors would be $8000$, which substantially exceeds the number seconds of walking for each individual. Second, these covariates are likely to be highly correlated with each other, which makes the design matrix very close to being rank deficient. Third, the fit depends strongly on the choice of number of knots and their placement. 

To address these problems we propose to use penalized splines, where a quadratic penalty is imposed on $\boldsymbol{\beta}$. More precisely, we maximize the penalized log likelihood $
 l(\boldsymbol{Y}; \boldsymbol{\beta})-\boldsymbol{\beta}^t\mathbf{S}_{\boldsymbol{\lambda}}\boldsymbol{\beta} \;$,
where $l(\boldsymbol{Y}; \boldsymbol{\beta})$ is the Bernoulli log likelihood, $\mathbf{S}_{\boldsymbol{\lambda}}$ is a block diagonal penalty matrix specific for the tri-variate Kronecker product of splines, and $\boldsymbol{\lambda}$ is a length three vector of smoothing parameters. There are many options for choosing the structure of the penalty, but here we use derivative-based penalties \cite{wood_p-splines_2017}. Such a model can be fit using the {\ttfamily mgcv::gam} function, as described below 

\noindent \begin{verbatim} gam(Y ~ te(D_i, S_i, U_i, by = lmat), family=binomial, method="REML") \end{verbatim}

While this code is easy to read, it requires careful manipulation of the accelerometry time series into the appropriate predictor matrices {\ttfamily D\_i}, {\ttfamily S\_i}, and {\ttfamily U\_i}. We provide a full description of this process in the supplementary material and the R code is available at \url{https://github.com/lilykoff/ml_walking_fingerprint}. Once these models are fit, one can estimate the probabilities $\widehat{p}_{ij}^{i_0}$ and study participants can be predicted using the same techniques described for image partitioning.

\section{Results/Applications}
 
\subsection{Prediction} 
\subsubsection{Train/test split}
For the IU data, $75$\% of the data from each subject were used for training and the remaining $25$\% were used for testing the models. The training and testing data were randomly sampled at the second level, so time ordering was not preserved; previous analyses \cite{koffman_fingerprinting_2023} showed that ignoring the ordering of the seconds does not have a large impact on predictive accuracy; however, the sampling was stratified by subject to preserve the original proportion of data from each individual. For the ZJU data, two separate prediction tasks were performed. In the first, only data from session 1 was used. As with the IU data, $75\%$ of the data from each subject were used for training and the remainder were used for testing. We henceforth refer to this task as ``ZJU S1." For the second task, data from session 1 were used to train the model, and data from session 2 were used for testing; this is referred to as ``ZJU S1S2" in the rest of the paper. The purpose of the second task was to investigate the performance of each model when using data collected at different time points. Note that, intuitively, it may be easier to predict walking of an individual within the same walking session than from two walking sessions that are weeks apart. Table~\ref{tab:train_test_seconds} summarizes the number of minutes used for training and testing for each data set and task. For example, the first row indicates that for the  the IU data, a median (IQR) of $367$ ($51.5$) seconds per subject are used for training while a median (IQR) of $122.5$ ($16.25$) seconds per subject are used for testing. 

\begin{table}[ht]

\centering 
\caption{Median and IQR of the number of seconds used across subjects for each data set and prediction task}
\label{tab:train_test_seconds}
\begin{tabular}{|l|ll|ll|}
\hline
\multicolumn{1}{|c|}{\multirow{2}{*}{Data and task}} & \multicolumn{2}{c|}{Median (s)}         & \multicolumn{2}{c|}{IQR (s)}            \\ \cline{2-5} 
\multicolumn{1}{|c|}{}                               & \multicolumn{1}{l|}{Training} & Testing & \multicolumn{1}{l|}{Training} & Testing \\ \hline\hline
IU                                                   & \multicolumn{1}{l|}{367}      & 122.5   & \multicolumn{1}{l|}{51.5}     & 16.25   \\ \hline
ZJU S1                                               & \multicolumn{1}{l|}{49}       & 17      & \multicolumn{1}{l|}{13.0}     & 5.00    \\ \hline
ZJU S1S2                                             & \multicolumn{1}{l|}{66}       & 65      & \multicolumn{1}{l|}{18}       & 15      \\ \hline
\end{tabular}
\end{table}

\subsubsection{Identification Rates}
To quantify how well individuals are predicted by their own walking data, we calculated the rank-$1$ (how often an individual was ranked as the most likely individual using their walking data) and rank-$5$ (how often an individual was ranked among the top five likeliest individuals) for each method, data set, and prediction task. By definition, rank-$5$ accuracy is always larger than or equal to rank-$1$ accuracy. The accuracy for each model and task are summarized in Table ~\ref{tab:accuracy_summary}.

For the IU data the rank-$1$ accuracy of the logistic and functional regression approaches is perfect, while that of machine learning is almost perfect, misclassifying just one study participant. Rank-$5$ accuracy for all methods is perfect in the IU data. This is not surprising given our previous results and the way the study was conducted. Indeed, the task here is to recognize individuals using around six minutes of training data from the same walking session, where the complete session consisted of one long walk. This provides enough data to populate the image (complete autocorrelation distribution), while the image remains relatively stable between the training and testing sets. In this setting the method for building predictors, not the specific prediction algorithm, does the heavy lifting.

For the ZJU S1 task, functional regression performs best, achieving $98$\% rank-1 accuracy and $100$\% rank-5 accuracy.  Logistic regression was a close second with  $93$\% rank-1 accuracy and $99$\% rank-5 accuracy. This is remarkable, as the original methods were not developed for training on short walking intervals; here we used less than one minute of training data per person. Machine learning had only $71$\% rank-1 accuracy, but a rank-5 accuracy ($97$\%) comparable to that of logistic and functional regression. 

Predicting session 2 using data from session 1 in the ZJU data (ZJU S1S2) is the most difficult task for all models. Machine learning ($54$\%)and functional regression ($53$\%) performed best in terms of rank-1 accuracy. While logistic regression had a lower rank-1 accuracy of $41$\%, its rank-5 accuracy ($75$\%) was comparable to that of machine learning ($76$\%) and functional regression ($69$\%). It is likely that the poor rank-1 accuracy of logistic regression is due to the large number of predictors used, which was not as well balanced by the reduced number of observations. Given the difficulty of the task and the small data sets, the performance of these methods is exceptional. As a side note, the number of correctly identified individuals under a permutation of labels would follow a Poisson distribution with mean $1$. That is, the number of correctly identified individuals at random is at most $3$ or $4$, irrespective of the number of individuals in the data set \cite{fingerprintingCaffo}.

\begin{table}[ht]

\caption{Summary of Accuracy}
\label{tab:accuracy_summary}
\centering
\begin{small}
\begin{tabular}[t]{|p{18mm}|ll|p{16mm}|p{16mm}|p{14mm}|p{14mm}|p{10mm}|}
\hline
Data and Task & Model & & Rank-1 Accuracy & Rank-5 Accuracy & Rank-1 Correct & Rank-5 Correct & Total (n)\\
\hline \hline 
\multirow{3}{*}{IU} & \multirow{2}{*}{Image partitioning} & Logistic & 1.00 & 1.00 & 32 & 32 & \multirow{3}{*}{32}\\
 & & ML & 0.97 & 1.00 & 31 & 32 & \\ \cline{2-7}

 & Functional regression & &  1.00 & 1.00 & 32 & 32 & \\
\hline \hline 
\multirow{3}{*}{ZJU S1}  & \multirow{2}{*}{Image partitioning} & Logistic & 0.93 & 0.99 & 140 & 151 & \multirow{3}{*}{153} \\
& & ML & 0.71 & 0.97 & 109 & 149 & \\
\cline{2-7}
& Functional regression & &  0.98 & 1.00 & 150 & 153 & \\
\hline \hline 
\multirow{3}{*}{ZJU S1S2} & \multirow{2}{*}{Image partitioning} & Logistic & 0.41 & 0.75 & 63 & 114 & \multirow{3}{*}{153} \\
 & & ML & 0.54 & 0.76 & 82 & 117 & \\
\cline{2-7}
 & Functional regression &  & 0.53 & 0.69 & 81 & 106 & \\
\hline 
\end{tabular}
\end{small}
\end{table}

\subsubsection{Sensitivity analysis to the number of seconds in the testing set}\label{subsubsec:classification}

\begin{figure}[!ht]
\centering
 \includegraphics[width=.75\textwidth]{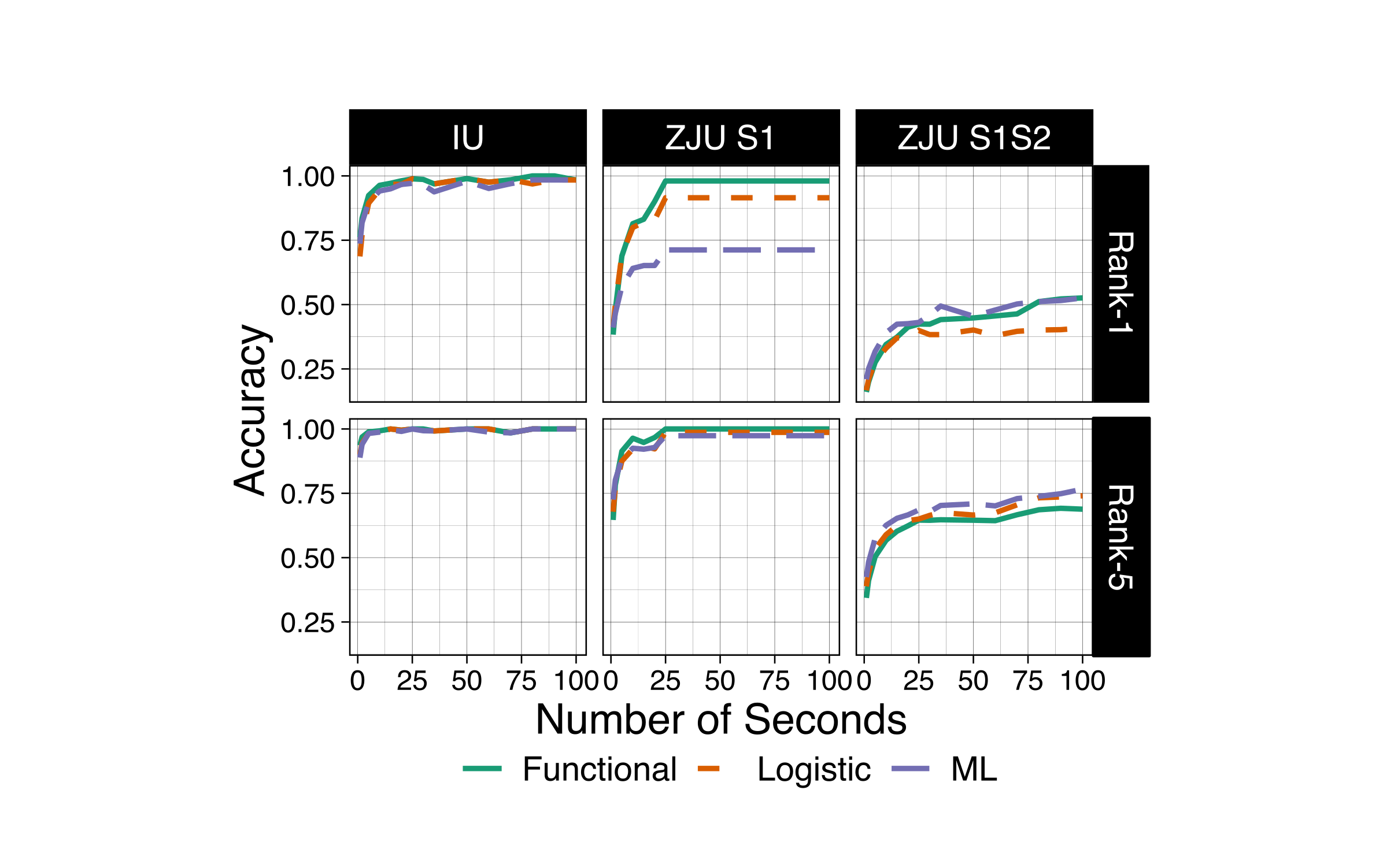}   
 \caption{Classification Metrics over Varying Number of Seconds in Testing Data. First row: rank-1 accuracies; Second row: rank-5 accuracies. Each column corresponds to different data and prediction tasks. The lines show how accuracy for each model changes as the number of seconds averaged over in the testing data is increased.}
 \label{fig:accuracies}
\end{figure}

So far, we have reported accuracy based on averaging over all seconds in the testing data to get a single prediction for each individual. However, we can average over shorter intervals to investigate how sensitive methods are to the number of seconds in the testing data. Figure~\ref{fig:accuracies} displays rank-1 and rank-5 accuracy for the three models, averaged over a varying number of seconds for the three prediction tasks. For example, lines in the upper-left hand panel of Figure~\ref{fig:accuracies} demonstrates how the rank-1 accuracy changes for the functional regression (green, solid line), logistic regression (orange, dashed line), and machine learning (purple, long-dashed line) as we average predictions in the test data over between $1$ second and $100$ seconds. As we average over more seconds in the testing data, the accuracy increases, because information is accumulated over multiple seconds and variability decreases. For the IU data, near-perfect rank-1 accuracy is achieved when averaging over at least $25$ seconds in the testing data for all models and perfect rank-5 accuracy is achieved when averaging over more than $1$ second. Essentially, this means that the top 5 predictions for each second in the testing data almost always contain the true subject. For the ZJU S1 task, averaging over increasing numbers of seconds in the testing data improves rank-1 and rank-5 accuracy for all models; improvements level off at $25$ seconds because few individuals have more than $25$ seconds of testing data present. The difference in the patterns in the IU data and ZJU S1 task likely arises from the smaller training data sets available in ZJU S1. The effect is that the empirical autocorrelation distribution is not completely filled in, but as more seconds are used for testing, the likelihood for recognizing the individual from multiple seconds of data increases. The final column shows that, for the ZJU S1S2 task, improvements in rank-1 and rank-5 accuracy level off at around $50$ seconds of testing data for all models. This is likely due to the fact that the walking of some individuals is so different between sessions 2 and 1, that they are difficult to identify even if they walk longer. 

\subsubsection{Machine Learning Models}\label{subsubsec:ML}
Several different machine learning models were considered for each individual. The model that achieved the highest AUC in five-fold cross-validation in the training data was selected as the model for each individual and then used to predict subject's identity on the testing data. Thus, different models were used for different subjects. In the IU data, the boosted tree was the best model (by AUC) for $12$ individuals, followed by the random forest ($9$), radial basis support vector machine (SVM) ($8$), polynomial basis SVM ($2$) and Bayesian additive regression tree (BART) ($1$). 
In the ZJU S1 data, the radial basis SVM ($43$), BART ($35$), and random forest ($24$) were the most accurate in cross-validation, but penalized regression ($15$), naive Bayes classifier ($7$), and logistic regression ($5$) were best for some individuals. Across all data and subjects, k-nearest neighbors, flexible discriminant analysis, and multiple additive regression splines (MARS) were never the best model.  

\subsection{Inference} \label{subsec:inference}
\subsubsection{Correlation and multiplicity adjustment (CMA)}
\begin{figure}[!htbp]
    \centering
    \begin{subfigure}[t]{0.45\textwidth}
        \centering
        \includegraphics[width=\linewidth]{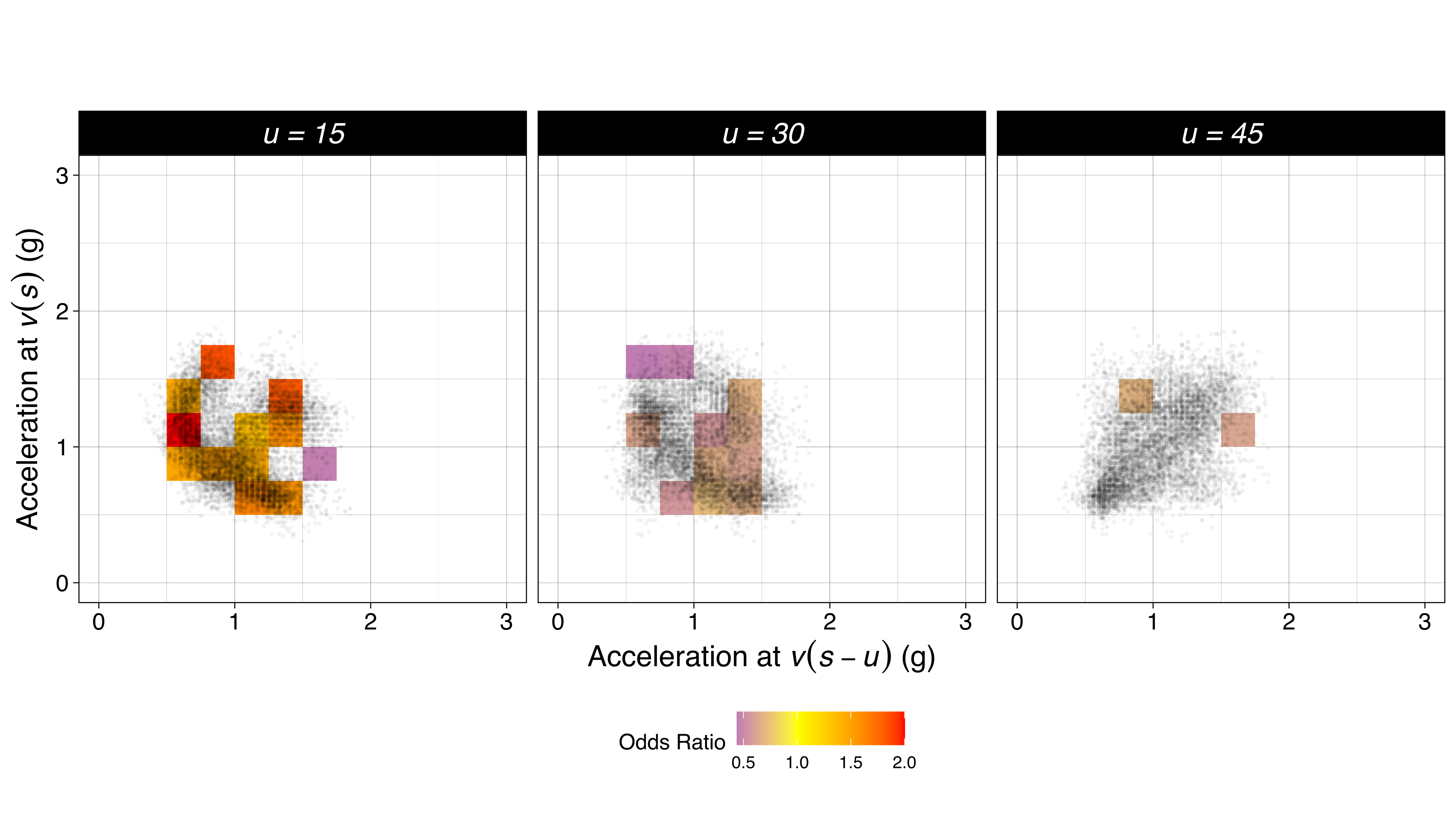} 
        \caption{Unadjusted} \label{fig:timing1}
    \end{subfigure}
    \hfill
    \begin{subfigure}[t]{0.45\textwidth}
        \centering
        \includegraphics[width=\linewidth]{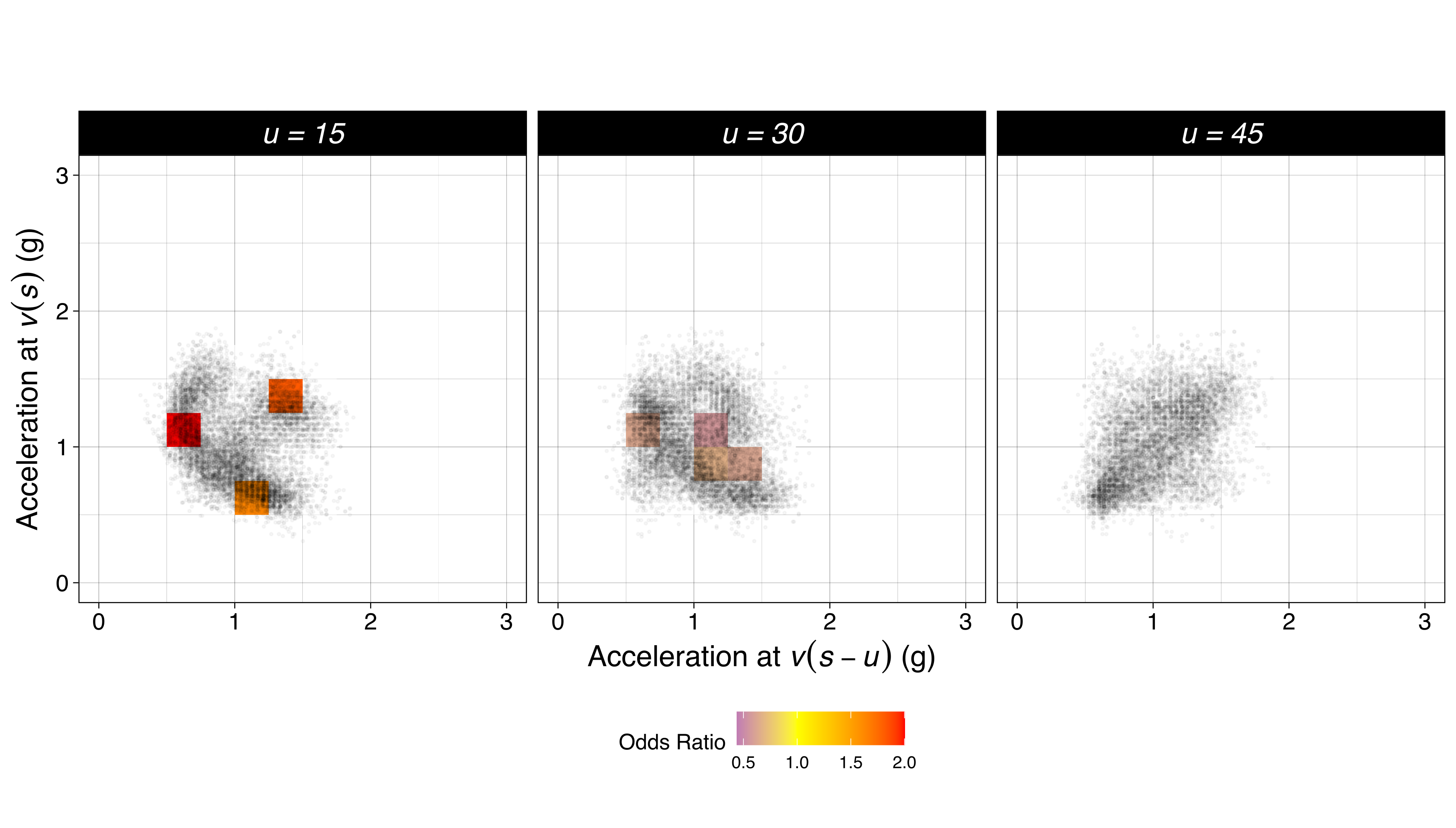} 
        \caption{Correlation and Multiplicity Adjusted} \label{fig:timing2}
    \end{subfigure}
    \label{fig:cma}
    \caption{Significant grid cells from image partitioning, Subject 143 ZJU Session 1. 
    Panel (a): grid cells that are significant in distinguishing subject 143 from the other subjects in the ZJU S1 task. Panel (b):  grid cells that are significant after adjusting for correlation and multiplicity.}
\end{figure}

Correlation and multiplicity adjusted (CMA) confidence intervals were calculated for each of the image-partitioning based logistic regression models to obtain estimates for the effect of points in each grid cell on the odds of a subjects' identity. To illustrate the use of CMA in prediction, we consider results from ZJU S1, predicting individuals within the same visit. Figure~\ref{fig:timing1} displays the grid cells identified from the logistic regression for subject $143$ using unadjusted (for correlation and multiplicity) confidence intervals and p-values. Figure~\ref{fig:timing2} displays the grid cells that remain significant after adjusting for  correlation and multiplicity of the tests. Each grid cell is colored according to its point estimate, where darker shades of red correspond to larger coefficients. For example, the dark orange grid cell located at $[1.25, 1.5], [1.25, 1.5]$ in the left-most ($15$ centisecond lag) column of panel (a) indicates that a pair of points $(v_{ij}(s) \in [1.25, 1.5), v_{ij}(s-15) \in [1.25, 1.5]$ increases the odds of being subject $143$. We would like to highlight that these findings are not obvious, would not be easy to identify by a human observer of the original data, and have immediate practical implications. After correlation and multiplicity adjustment, in this example, only seven grid cells are significant, compared to $25$ in the unadjusted setting. For the IU participants, the number of significant cells after adjustment ranges from $0$ to $22$; in the ZJU participants it ranges from $0$ to $61$. The reduced number of cells allows us to focus on the areas of the fingerprint that are most important for distinguishing the individuals from the others.

\subsubsection{Walking fingerprints}

\begin{figure}[!htbp]
    \centering
    \begin{subfigure}[t]{0.75\textwidth}
        \centering
        \includegraphics[width=.95\linewidth]{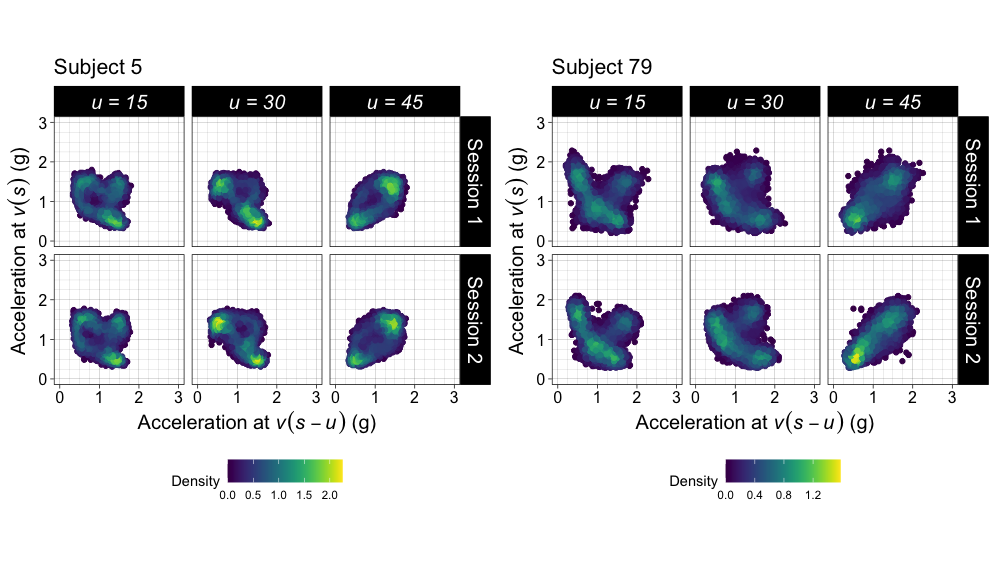} 
        \caption{Well Predicted Subjects} \label{fig:timing1_rep}
    \end{subfigure}
    \hfill
    \begin{subfigure}[b]{0.75\textwidth}
        \centering
        \includegraphics[width=.95\linewidth]{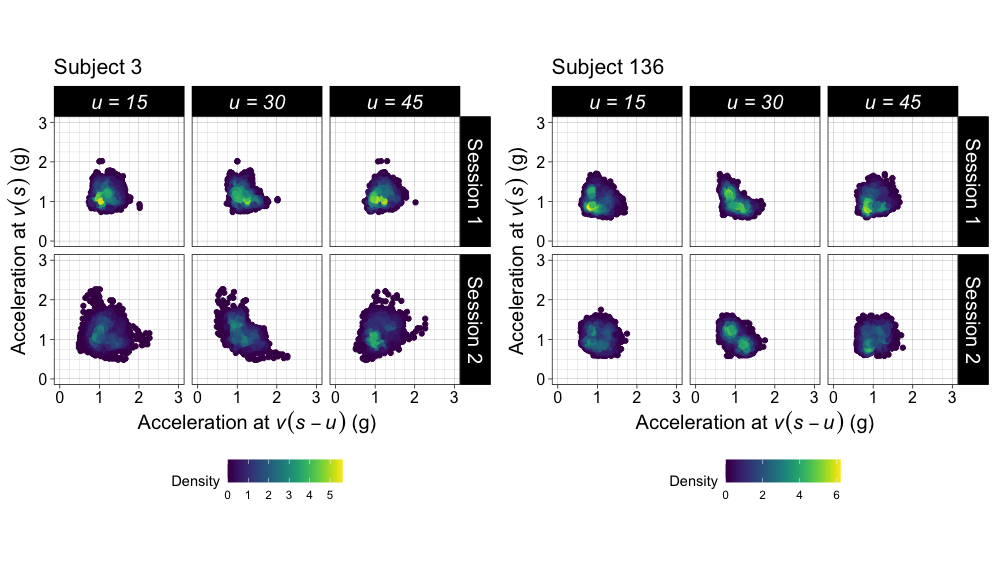} 
        \caption{Poorly Predicted Subjects} \label{fig:timing2_rep}
    \end{subfigure}
    \label{fig:fingerprints}
    \caption{Comparison of Data from Well and Poorly Predicted Subjects\\
    \footnotesize Panel (a) demonstrates a subset of the autocorrelation distribution for subject 5 (left) and subject 79 (right) in session 1 (top row) and session 2 (bottom row). The images are similar between sessions and these individuals were correctly identified in session 2 from their session 1 data. Panel (b) shows the same subset of the autocorrelation distribution for subject 3 (left) and subject 136 (right) in session 1 (top row) and session 2 (bottom row). The images do not look similar, and hence these individuals were not correctly predicted in the ZJU S1S2 task.}
\end{figure}

Visualization of a subset of the empirical autocorrelation distribution provides further insight into how the methods work.  Figure~\ref{fig:timing1_rep} displays the walking fingerprint for session 1 and session 2 for individuals who were correctly predicted between sessions, and Figure ~\ref{fig:timing2_rep} displays images for sessions 1 and 2 for those who were not correctly predicted. Clearly, those who were correctly predicted had much more consistent patterns between the two sessions.  This supports the idea that the methods described in this paper are homing in on important characteristics of walking and are sensitive to within-person changes in walking.

\section{Discussion}

We have proposed two frameworks for identifying individuals from their walking pattern derived from wearable accelerometers. In both frameworks, the raw time series is first transformed into the complete empirical autocorrelation distribution, which can be thought of as a three dimensional image. In the first framework, scalar predictors are obtained as summaries of this image and used in one versus the rest classification using logistic regression or machine learning models. We refer to this approach as image partitioning because predictors are derived by summing over areas of a 3D image. In the second framework, the entire autocorrelation distribution is used in a trivariate functional regression model. The two frameworks were deployed in two separate data sets and three different scenarios using publicly available accelerometry data. When there is at least $5$ minutes of data observed for each individual and we predict within the same session, as in the IU data, the performance of the two frameworks is similar: the functional regression slightly outperforms image partitioning approaches, but all achieve greater than $95$\% accuracy. In the ZJU data when predicting within the same session, functional regression performs best, followed by the logistic regression and machine learning. For the difficult task of predicting the identity of individuals in a second session at least a week after the first, the functional regression and machine learning perform best, followed by logistic regression. The results are a large improvement over our previous work.

The intuition behind the two frameworks is the same. The way people walk is semi-unique, but walking is inherently a repetitive movement. For a regression-based approach to accurately predict individuals, it must leverage the repetitive nature of the data. One way to harness this repetition is to look at autocorrelation, and both the functional regression and image partitioning approaches use autocorrelation. The functional approach uses all pairwise autocorrelations into a regression model. The image partitioning approach selects a subset of pairwise autocorrelations and then creates predictors by counting the number of these pairs that fall in different categories.

While the intuition behind the frameworks is the same, the implementation and results differ. The functional approach appears to perform better in smaller data sets and with truly out of sample prediction; however, it is computationally more expensive and may be difficult to scale in large data sets. The image partitioning approach is less computationally intensive and allows for inference by identifying areas that distinguish each individual, and generates a unique walking ``fingerprint" \cite{koffman_fingerprinting_2023}. However, image partitioning seems to perform worse in smaller data sets and with out of sample prediction. Furthermore, it requires decisions about which lags and how many lags to use and requires variable selection to avoid overfitting. In even larger data sets, the logistic regression approach may be preferable, as it takes only minutes to run compared to several hours for the machine learning and functional regression methods; the functional regression also requires substantial memory.

Few existing methods for identifying individuals from their walking pattern report recognition rate or accuracy of their algorithms. However, our model outperforms Gafurov et al. ($86.3$\% accuracy in a study of $50$ individuals)  \cite{gafurov2007gait} and is comparable to Pan et al ($96.7$\% accuracy in a study of $30$ individuals)\cite{pan_accelerometer-based_2009} and Zhang et al ($95.8$\% accuracy in the ZJU data used here) \cite{zhang_accelerometer-based_2015}. However, both Pan et. al. and Zhang et. al. used data from five locations (wrist, upper arm, hip, knee, and ankle), while we use only data from the wrist. When using just data from the wrist, Zhang et. al. achieved a recognition rate of $56.4$\%.

The prediction was highly affected by the difficulty of the task. Indeed, predicting individuals from a walking session weeks before the current walking session was more difficult. This is likely because the time gap between the collection of the first and second session presents many challenges. In spite of these challenges, methods performed very well.

However, there are still limitations to this data and study. The data in both applications were collected in semi-controlled environments and labeled as walking. It remains unclear how these methods would perform on free-living data and on more heterogeneous populations. 

Finally, we promised that we would solve the puzzle from Figure~\ref{fig:problem_description} at the end of the paper. Data from subjects 43 and 118 in the ZJU data are displayed in the second and fourth row of the right-hand panel, respectively, while the data in the first and third rows belong to subjects 28 and 80. 

\medskip

\subsection*{Funding acknowledgments}
This work was supported by the National Institutes of Health under Grant R01NS060910 and Grant R01AG075883. 
\subsection*{Conflict of Interest Disclosure Statement}
Ciprian Crainiceanu is consulting for Bayer and Johnson and Johnson on methods development for wearable and implantable technologies. The details of these contracts are disclosed through the Johns Hopkins University eDisclose system. The research presented here is not related to and was not supported by this consulting work.
\bigskip

{\small
\bibliography{references}}

\clearpage

\section*{Supplementary Material}
\subsection*{Preparing raw accelerometry data for functional regression}
\noindent We use the same notation as in section 3.1 and 3.2 of the manuscript: $v_{ij}(s)$ is the vector magnitude of acceleration in gravitational units (g) at second $j$ for subject $i$. In section 3.4, we note that the trivariate functional regression model can be fit with the single line: 

\noindent \begin{verbatim} gam(Y ~ te(D_i, S_i, U_i, by = lmat), family=binomial, method="REML") \end{verbatim}

\noindent Here we describe how to manipulate the raw time series data into the format such that this line of code can be deployed.

 \bigskip 
 
For the functional regression approach, we first obtain the empirical complete autocorrelation distribution for each subject, then manipulate the distribution into three matrices for each subject, which are used in the functional regression. Recall that $S$ denotes the interval length used to obtain the autocorrelation distribution, in our case we let $S = 100$ centiseconds, and $J_i$ is the number of seconds observed for individual $i$. Then $\mathbf{D}_i, \mathbf{S}_i, \mathbf{U}_i \in \mathbb{R}^{J_i\times S(S-1)/2}$. Conceptually, one can think of $\mathbf{U}_i$ as a matrix of lags, $\mathbf{S}_i$ as a matrix of accelerations, and $\mathbf{D}_i$ as a matrix of lagged accelerations. 
Let $\mathbf{1}_n \in \mathbb{R}^{1\times n}$ be the row vector of all ones. Then:

\begin{align*}
   \mathbf{D}_i = 
   \begin{bmatrix}
   v_{i1}(1) \mathbf{1}_{99} &  v_{i1}(2) \mathbf{1}_{98} &\dots &  v_{i1}(99) \mathbf{1}_{1}  \\
    v_{i2}(1) \mathbf{1}_{99} &  v_{i2}(2) \mathbf{1}_{98} &\dots &  v_{i2}(99) \mathbf{1}_{1}  \\
    \vdots & \vdots & \ddots &\vdots \\
    v_{iJ_i}(1) \mathbf{1}_{99} &  v_{iJ_i}(2) \mathbf{1}_{98} &\dots &  v_{iJ_i}(99) \mathbf{1}_{1} 
    \end{bmatrix}
\end{align*}

\noindent Let $\mathbf{v}_{ij}(k:l) \in \mathbb{R}^{1\times (l-k+1)}$ be a row vector of observations for subject $i$ at second $j$ for centiseconds $k$ through $l$, where $l \geq k$. Then: 
\begin{align*}
    \mathbf{S}_i = \begin{bmatrix}
        \mathbf{v}_{i1}(2:100) & \mathbf{v}_{i1}(3:100) & \dots & \mathbf{v}_{i1}(100:100)\\ 
         \mathbf{v}_{i2}(2:100) & \mathbf{v}_{i2}(3:100) & \dots & \mathbf{v}_{i2}(100:100)\\ 
        \vdots & \vdots & \ddots & \vdots \\ 
       \mathbf{v}_{iJ_i}(2:100) & \mathbf{v}_{iJ_i}(3:100) & \dots & \mathbf{v}_{iJ_i}(100:100)\\ 
    \end{bmatrix}
\end{align*}
Finally, let $a:b$ for $a,b\in \mathbb{Z}, a < b$ denote the sequence  $a,a+1, a+2, \dots,  b-1, b $. Then: 
\begin{align*}
   \mathbf{U}_i  = \mathbf{1}_{J_i\times 1}
   \begin{bmatrix}
       1:99 & 1:98 & 1:97 & \dots & 1:2 & 1 
   \end{bmatrix}
   \end{align*}

\subsubsection*{Toy Example}

Suppose we have a situation with interval length $S = 4$, and we observe just two intervals per subject, i.e. $J_i = 2 \forall i$.  Then the raw data is $v_{ij}(k) \text{ for } j = 1,2; k = 1, 2, 3, 4$. Then the matrices for subject $i$ are as follows: 

\begin{align*}
    \mathbf{D}_i = 
   \begin{bmatrix}
   v_{i 1}(1)  & v_{i1}(1)& v_{i 1}(1)  & v_{i 1}(2) & v_{i 1}(2) & v_{i 1}(3) \\
      v_{i 2}(1)  & v_{i 2}(1)& v_{i 2}(1)  & v_{i 2}(2) & v_{i 2}(2) & v_{i 2}(3) \\
    \end{bmatrix}
\end{align*}

\begin{align*}
    \mathbf{S}_i = 
   \begin{bmatrix}
   v_{i 1}(2)  & v_{i 1}(3)& v_{i 1}(4)  & v_{i 1}(3) & v_{i 1}(4) & v_{i 1}(4) \\
      v_{i 2}(2)  & v_{i 2}(3)& v_{i 2}(4)  & v_{i 2}(3) & v_{i 2}(4) & v_{i 2}(4) \\
    \end{bmatrix}
\end{align*}

\begin{align*}
    \mathbf{U}_i = 
   \begin{bmatrix}
   1 & 2 & 3 & 1 & 2 & 1\\
     1 & 2 & 3 & 1 & 2 & 1\\
    \end{bmatrix}
\end{align*}

\subsubsection*{Fitting the model}
The remaining arguments in the {\ttfamily mgcv::gam} syntax are $\mathbf{Y}$ and {\ttfamily lmat}. $\mathbf{Y}$ is a column vector of ones and zeroes, where $Y_{ij}^{i_0}=1$ if $i=i_0$ and $0$. 
{\ttfamily lmat} is a matrix of the same dimension as $\mathbf{D}_i, \mathbf{S}_i, \text{and } \mathbf{U}_i$ and contains the weights of the linear functionals of the smooth terms; in our case we use equal weights, so the $(i, j)$ entry of {\ttfamily lmat} is $\frac{1}{S(S-1)/2}$ for all $i$ and $j$. 

The call to {\ttfamily te()} specifies that we want to form a tensor product smooth of the variables supplied as the first unnamed arguments to the function. Adding {\ttfamily method="REML"} to the function call specifies that smoothing parameter selection is done using restricted maximum likelihood. Because the basis is constructed using a tensor product smooth of three variables, there are three smoothing parameters which must be selected by the model.

The subject-specific matrices and {\ttfamily lmat} may be put together in a data frame using the \texttt{AsIs} function in R, and then the functional regression can be fit as described in the manuscript and code 
\url{https://github.com/lilykoff/ml_walking_fingerprint}

\end{document}